\documentclass[aps,prb,twocolumn,groupedaddress,floatfix]{revtex4-1}

\usepackage{graphicx}
\usepackage{dcolumn}
\usepackage{bm}

\begin{document}

\title{Electrical control of phonon mediated spin relaxation rate in  semiconductor quantum dots: the Rashba vs the Dresselhaus spin-orbit couplings}
\author{Sanjay Prabhakar,$^1$  Roderick Melnik$^{1,2}$ and Luis L. Bonilla$^2$ }
\affiliation{
$^1$M\,$^2$NeT Laboratory, Wilfrid Laurier University, Waterloo, ON, N2L 3C5 Canada\\
$^2$Gregorio Millan Institute, Universidad Carlos III de Madrid, 28911, Leganes, Spain
}

\date{May 15, 2013}

\begin{abstract}
In symmetric quantum dots (QDs), it is well known that  the spin-hot spot (i.e., the cusp-like structure due to the presence of degeneracy near the level  or anticrossing point) is present for the pure Rashba case but is absent for the pure Dresselhaus case [Phys. Rev. Lett. 95, 076805 (2005)]. Since  the Dresselhaus spin-orbit coupling dominates over the Rashba spin-orbit coupling in GaAs and GaSb QDs, it is important to find the exact location of  the spin-hot spot or the cusp-like structure   even for the pure Dresselhaus case.  In this paper, for the first time, we present  analytical and numerical results that show that the spin-hot spot   can also  be seen for the pure Dresselhaus spin-orbit coupling case by inducing large anisotropy through external gates.  At or nearby the spin-hot spot, the spin transition rate enhances and  the decoherence time reduces by several orders of magnitude compared to  the case with no spin-hot spot. Thus one should avoid such locations when designing   QD spin based transistors for the possible implementation in quantum logic gates, solid state quantum computing and quantum information processing. It is also possible to extract the exact experimental data (Phys. Rev. Lett. 100, 046803 (2008)) for the phonon mediated spin-flip rates  from our developed theoretical model.

\end{abstract}



\maketitle

\section{Introduction}

Manipulation of a single electron spin with the application of gate controlled electric fields in a confined semiconductor QDs is a promising way for developing spin based quantum logic gates, spin memory devices for various quantum information processing applications.~\cite{loss98,awschalom02,hanson05, kroutvar04,elzerman04,glazov10,climente07,pietilainen06,chakraborty05,nowak11a,nowak09,nowak10} Sufficiently short gate operation time combined with long decoherence time is  one of the requirements  for  quantum computing.~\cite{golovach04,wang12,loss98} When a qubit is operated on by a classical bit, then its decay time is  given by a spin relaxation time which is  also supposed to be longer than the minimum time required to execute one quantum  gate operation.~\cite{golovach04,awschalom02,amasha08,bandyopadhyay00,fujisawa01}
Long spin relaxations  have been measured experimentally in both  symmetric and  asymmetric QDs.~\cite{amasha08,elzerman04,kroutvar04}
Balocchi et. al.~\cite{balocchi11} have recently measured larger  spin relaxation times ($\mathrm{30~ns}$) in GaAs QDs. More specifically, both isotropic and anisotropic  spin relaxations can be tuned with spin orbit coupling by choosing the growth direction parallel to the crystallographic axis [001], [110] and [111] of III-V zinc blend semiconductor QDs.~\cite{glazov10,balocchi11,griesbeck12,olendski07}   In addition to the lengthening spin coherence time, the electric field tuning of spin relaxation forms the basis for turning the spin current on and off in some spin transistor proposals that can help to initialize electron spin based quantum computers.~\cite{bandyopadhyay00,flatte11} These experimental studies confirm
that the manipulation of spin-flip rates mediated by phonons due to spin-orbit coupling is an important ingredient for the design of robust spintronics logic devices.
The spin-orbit coupling is mainly dominated by the Rashba~\cite{bychkov84} and the linear Dresselhaus~\cite{dresselhaus55} terms in III-V semiconductor  QDs.~\cite{prabhakar09,prabhakar10,sousa03,prabhakar12,folk01,fujisawa01,fujisawa02,hanson03,pryor06,pryor07,nowak11} The Rashba spin-orbit coupling arises from structural inversion asymmetry along the growth direction while the Dresselhaus spin-orbit coupling arises from the bulk inversion asymmetry of the crystal lattice.~\cite{bychkov84,dresselhaus55}

In Ref.~\onlinecite{bulaev05,bulaev05a}, the authors report that the cusp-like structure in the phonon mediated spin transition rate can be  seen  for the pure Rashba case. For the pure Dresselhaus case, the spin transition rate is a monotonous function of the magnetic fields and QDs radii. Since  the Dresselhaus spin-orbit coupling dominates over the Rashba spin-orbit coupling in some materials such as GaAs and GaSb QDs,~\cite{prabhakar10} it is important to find  the exact location of spin-hot spot or the cusp-like structure   even for the pure Dresselhaus case. The cusp like structure implies shorter spin relaxation and decoherence time which is hazardous for spin based applications such as quantum logic gates, solid state quantum computing and quantum information processing. For these applications, the spin-hot spot in the phonon mediated spin relaxation rate is something  to avoid  during the design of  QD spin based transistors.  Very recently, the authors in Ref.~\onlinecite{yang12} measured the spin-hot spot in the phonon mediated spin relaxation rate  in Silicon QDs with the application of  tuning very weak spin orbit coupling  when Zeeman energy and valley splittings induce degeneracy. At the spin-hot spot in Silicon QDs, the dramatic rate enhancement decreases the decoherence time which is not supposed   to be the ideal location for the qubit operation.~\cite{weina13,koh12,zhan12,prance12} In this paper, we obtain new analytical and numerical results for the behavior of the spin relaxation rate in anisotropic III-V semiconductor QDs. For the first time, we show   the spin-hot spot  in the phonon mediated spin transition rate can be seen for the pure Dresselhaus case by creating large anisotropy through external gates. Note that such location (spin-hot spot) is hazardous for quantum computing and quantum information processing, and  must therefore  be avoided during the design of spin based transistors.

\begin{figure}
\includegraphics[width=8.5cm,height=6cm]{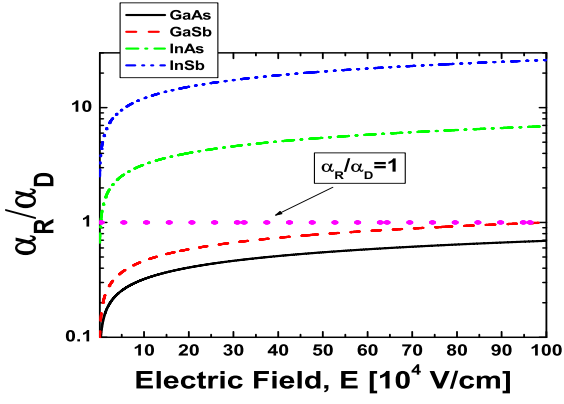}
\caption{\label{fig1}
(Color online) Interplay between Rashba-Dresselhaus spin-orbit coupling vs the applied electric field along the z-direction. The Rashba spin-orbit coupling is seen to dominate in InAs and InSb QDs whereas the Dresselhaus spin-orbit coupling is seen to dominate in GaAs and GaSb QDs.
}
\end{figure}
\begin{figure}
\includegraphics[width=8.5cm,height=6cm]{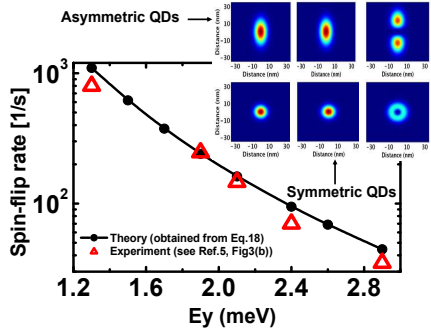}
\caption{\label{fig2}
(Color online) Relaxation rate vs anisotropy in QDs. We choose   $B=3~T$, $\ell_0=10~nm$, $\lambda_R=\lambda_D = 1.7~\mu m $ and $a=5$. Here we define  $\lambda_R=\hbar^2/m\alpha_R$, $\lambda_D=\hbar^2/m\alpha_D$, $E_x=\hbar\omega_0\sqrt a$ and $E_y=\hbar\omega_0\sqrt b$. The choice of these parameters  mimics the experimentally reported values in Ref.~\onlinecite{amasha08}. It can be seen that the theoretically obtained spin relaxation rate is in excellent agreement with the experimentally reported values in Ref.~\onlinecite{amasha08}. For symmetric QDs, (lower panel, inset plot), we chose $a=b=5$.
}
\end{figure}
\begin{figure*}
\includegraphics[width=18cm,height=10cm]{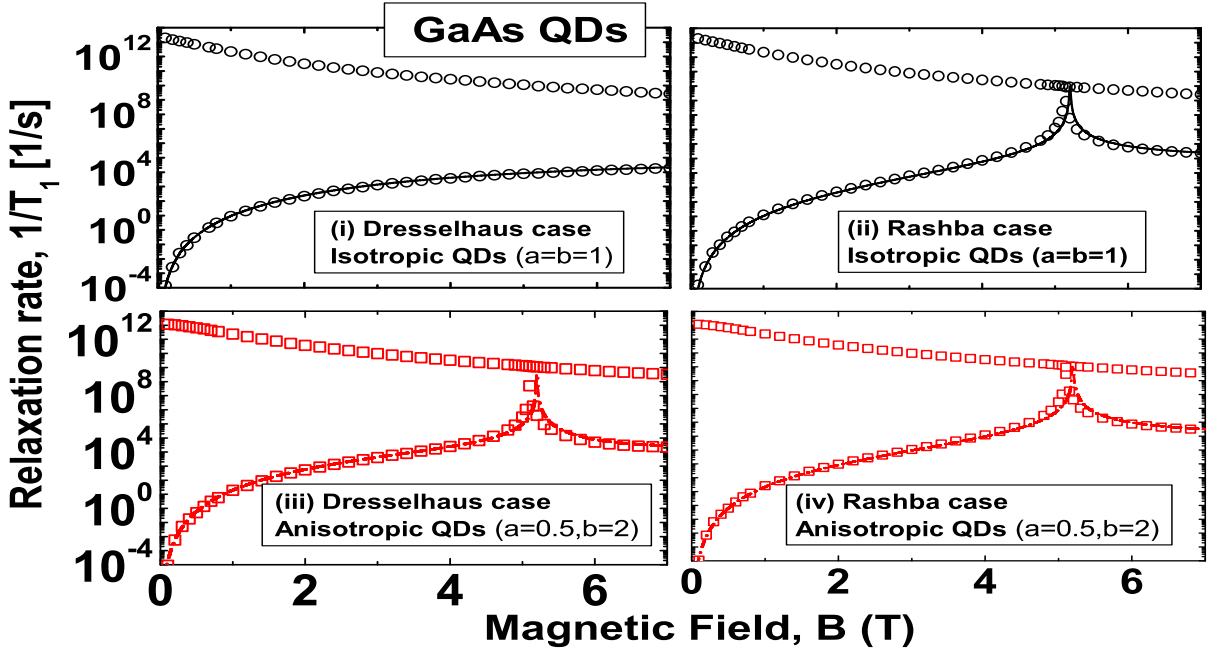}
\caption{\label{fig3}
(Color online)
Contributions of the Rashba and the Dresselhaus spin-orbit couplings to the phonon induced spin-flip rate
as a function of magnetic fields. Material constants are chosen the  same as in Fig.~\ref{fig1}, but  $\hbar \omega_0=1.1~meV$   and $\lambda_R=\lambda_D = 8 \mu m $. Solid lines (blue) are obtained from Eq.~\ref{1-T1-a}. Open circles and squares are obtained numerically from Eq.~\ref{1-T1} by an exact diagonalization scheme implemented via Finite Element Method.~\cite{comsol}   Notice that  a cusp-like structure can be seen for the pure Dresselhaus case in asymmetric QDs (Fig.~\ref{fig2} (iii), $a\neq b$), but not for symmetric QDs (Fig.~\ref{fig2} (i), $a=b$). Also, the spin-flip rate vanishes like $B^5$ (see Eq.~\ref{1-T1-1}). Fig.~\ref{fig2}(i) is  Loss et. al. proposal for symmetric QDs (see Ref.~\onlinecite{bulaev05}). Fig.~\ref{fig2}(iii) is our proposal for asymmetric QDs. \emph{We also expect a similar cusp-like structure  for the pure Dresselhaus case with heavy holes in asymmetric QDs  which is different from Ref.~\onlinecite{bulaev05}.}
}
\end{figure*}
\begin{figure*}
\includegraphics[width=18cm,height=10cm]{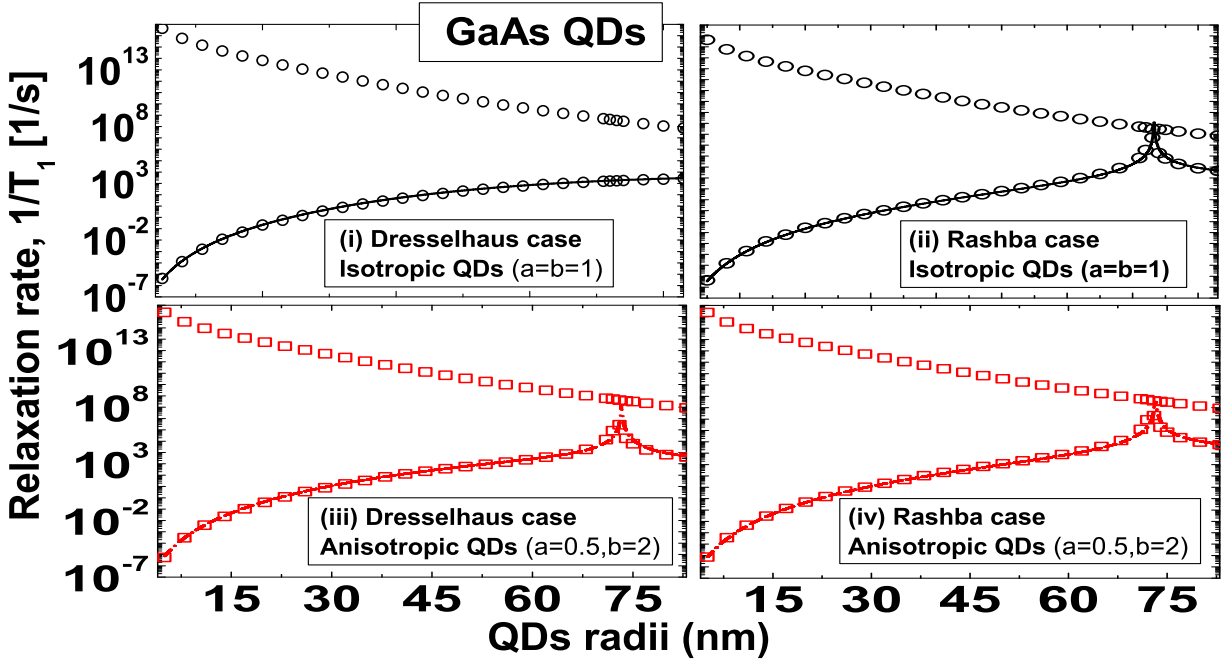}
\caption{\label{fig4}
(Color online) Same as Fig.~\ref{fig2} but $1/T_1$ vs $\ell_0$.
Here we chose, $B=1 T$. Again, notice the   cusp-like structure can only be seen for the pure Dresselhaus case in asymmetric QDs (Fig.~\ref{fig3} (iii), $a\neq b$) but not for symmetric QDs (Fig.~\ref{fig3} (i), $a=b$). Also, the spin-flip rate vanishes like $\ell_0^8$ (see Eq.~\ref{1-T1-1}).
}
\end{figure*}
\begin{figure*}
\includegraphics[width=18cm,height=12cm]{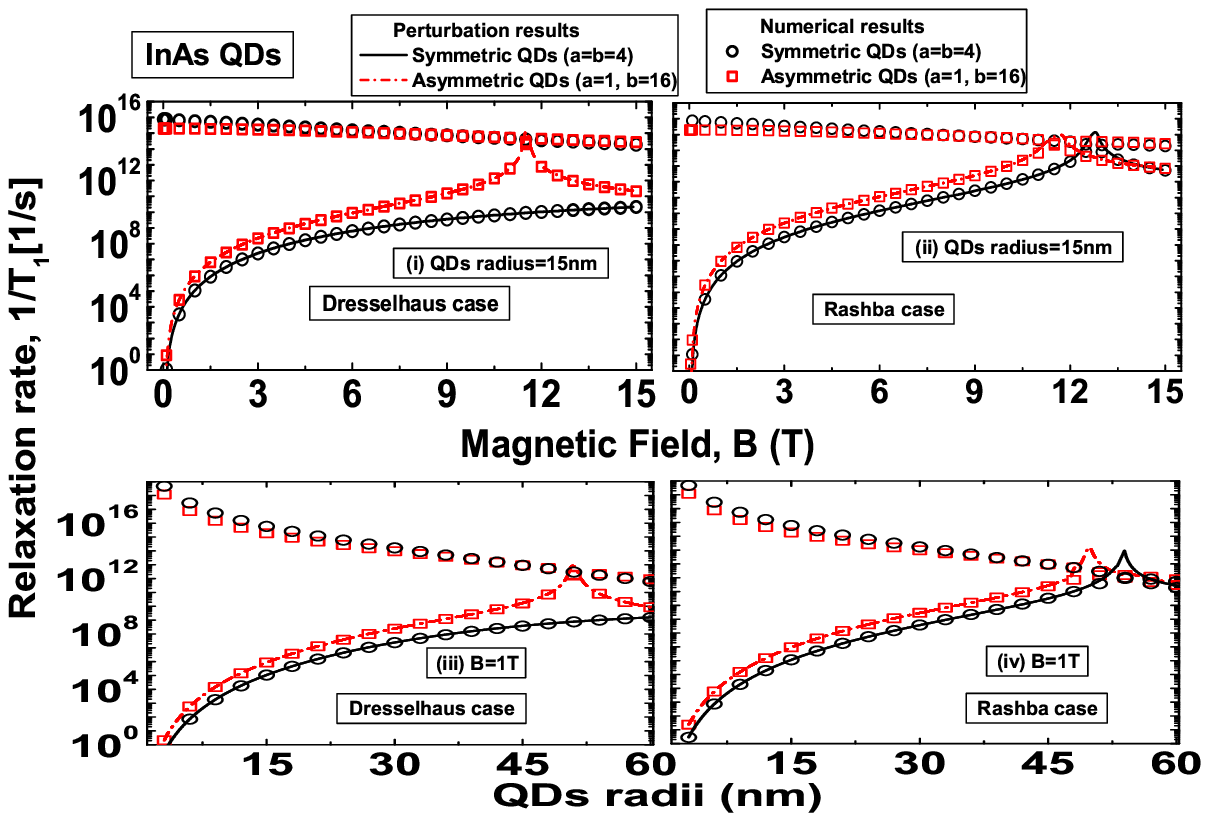}
\caption{\label{fig5}
(Color online) Same as Figs.~\ref{fig2} and~\ref{fig3}  but for InAs QDs. We chose  $E=10^5~V/cm$.
Again, notice the   cusp-like structure can also be seen for the pure Dresselhaus case in asymmetric QDs $(a\neq b)$, but not for symmetric QDs  $(a=b$).
}
\end{figure*}
\begin{figure*}
\includegraphics[width=18cm,height=12cm]{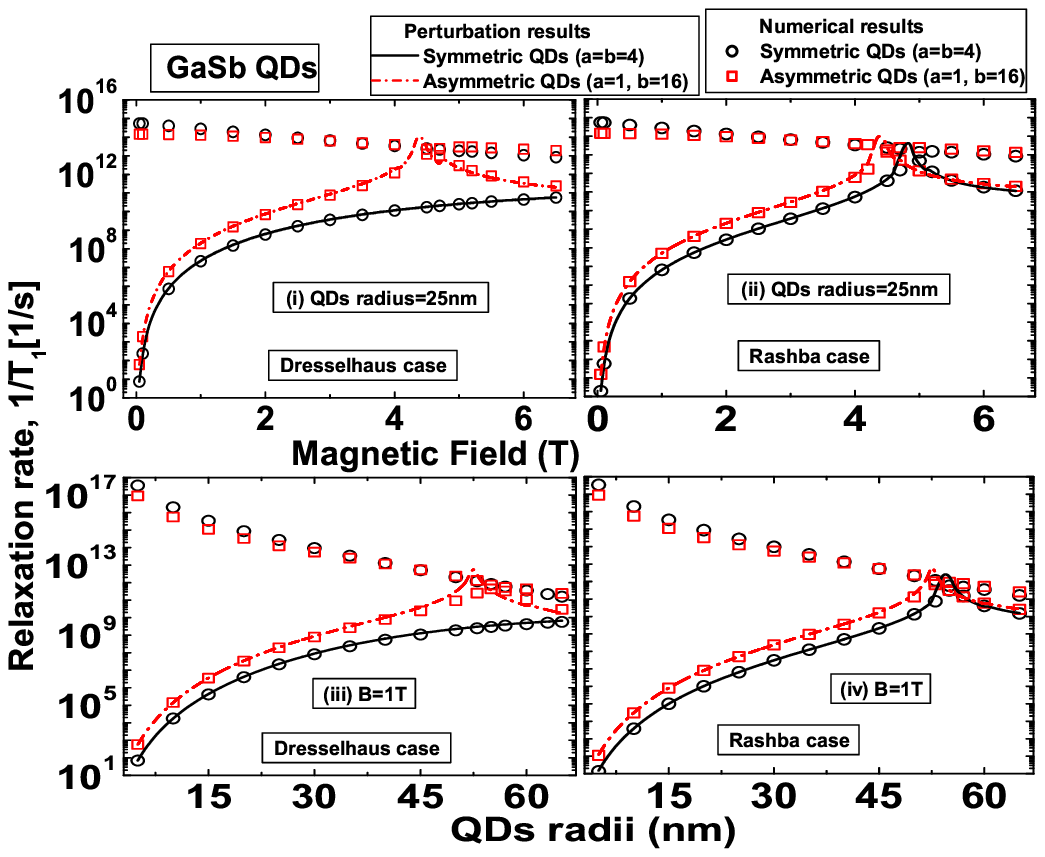}
\caption{\label{fig6}
(Color online) Same as Fig.~\ref{fig2} and~\ref{fig3},  but for GaSb QDs. We chose  $E=10^5~V/cm$.
Again, notice the   cusp-like structure can also be seen for the pure Dresselhaus case in asymmetric QDs $(a\neq b)$, but not for symmetric QDs  $(a=b$).
}
\end{figure*}
\begin{figure*}
\includegraphics[width=18cm,height=12cm]{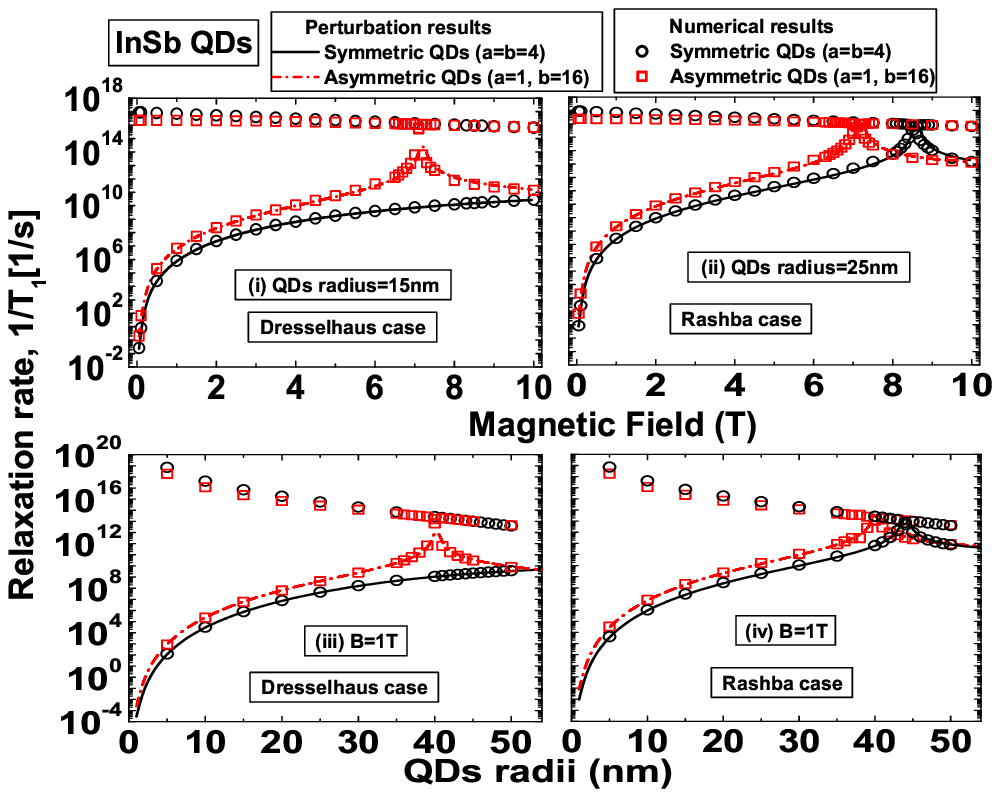}
\caption{\label{fig7}
(Color online) Same as Fig.~\ref{fig2} and~\ref{fig3}  but for InSb QDs. We chose  $E=10^4~V/cm$.
Again, notice the   cusp-like structure can also be seen for the pure Dresselhaus case in asymmetric QDs $(a\neq b)$, but not for symmetric QDs  $(a=b$).
}
\end{figure*}

The paper is organized as follows. In section~\ref{theoretical-model}, we develop a theoretical model for anisotropic spin relaxation mediated by piezo-phonons that will allow us to investigate the interplay between the Rashba and the linear Dresselhaus spin-orbit couplings  in QDs. In section~\ref{computational-method}, we provide  details of the diagonalization technique used for finding the energy spectrum and the matrix elements of the phonon mediated spin transition rate in QDs.   In section~\ref{results-and-discussions}, we plot both isotropic and anisotropic spin relaxation rates vs. magnetic fields and QDs radii for the pure Rashba and the pure Dresselhaus case in III-V semiconductor materials of zinc blend structure such as GaAs, GaSb, InAs and InSb.  Finally, in section~\ref{conclusion}, we summarize our results.

\section{Theoretical Model}\label{theoretical-model}

We consider 2D anisotropic  semiconductor QDs in the presence of a magnetic field along the growth
direction. The total Hamiltonian of an electron in anisotropic QDs including spin-orbit interactions can be
written as~\cite{sousa03,khaetskii00,bulaev05} $H = H_{xy}  +  H_{so}$,
where $H_{so}=H_R+H_D$  is the Hamiltonian associated with the Rashba-Dresselhaus spin-orbit couplings and $H_{xy}$ is the Hamiltonian of the electron in anisotropic QDs. $H_{xy}$ can be written as
\begin{equation}
H_{xy} = {\frac {\vec{P}^2}{2m}} + {\frac{1}{2}} m \omega_o^2
(a x^2 + b y^2) + {\frac 1 2} g_o \mu_B \sigma_z B,
\label{hxy}
\end{equation}
where $\vec{P} = \vec{p} + e \vec{A}$ is  the kinetic momentum operator, $\vec{p} = -i\hbar (\partial_x,\partial_y,0)$ is the canonical momentum operator, $\vec{A}=B \left(-y \sqrt b ,x\sqrt a,0\right)/\left(\sqrt a+\sqrt b\right)$  is the vector potential in the asymmetric gauge, $m$ is the effective mass,  $\mu_B$ is the Bohr magneton, $\vec{\sigma}=\left(\sigma_x,\sigma_y,\sigma_z\right)$ are the Pauli spin matrices, $g_0$ is the bulk g-factor,  $\omega_0=\hbar/(m\ell_0^2)$ is the parabolic confining potential and $\ell_0$ is the radius of the QDs.
The  energy spectrum of $H_{xy}$ can be written as~\cite{sousa03,prabhakar11}
\begin{equation}
\varepsilon^0_{n_+,n_-,\pm}=\left(n_++n_-+1\right)\hbar\omega_++\left(n_+-n_-\right)\hbar\omega_-\pm \frac{\Delta}{2}, \label{epsilon-a}\\
\end{equation}
where    $\omega_{\pm}=\frac{1}{2}\left[\omega_c^2+\omega_0^2\left(\sqrt a\pm\sqrt b\right)^2\right]^{1/2}$, $\Delta=g_0\mu_B B$ and $n_{\pm}$ are the eigenvalues of the Fock-Darwin number operators $a_{\pm}^\dagger a_{\pm}$. Here,  $a_{\pm}$ and $a_{\pm}^\dagger$ are usual annihilation  and creation  operators. Also, we label the  Fock-Darwin states as $|n_+, n_-,\pm\rangle$   with $\pm$ being the eigenvalues of the Pauli spin matrix along z-direction.~\cite{sousa03,prabhakar09}

Finally, $H_{so}$ can be written as~\cite{sousa03}
\begin{equation}
H_{so} =\frac{\alpha_R}{\hbar}\left(\sigma_x P_y - \sigma_y P_x\right)+ \frac{\alpha_D}{\hbar}\left(-\sigma_x P_x + \sigma_y P_y\right),\label{rashba-dresselhaus}
\end{equation}
where
\begin{equation}
\alpha_R=\gamma_ReE,~~\alpha_D=0.78\gamma_D\left(\frac{2meE}{\hbar^2}\right)^{2/3}.\label{alpha-R}
\end{equation}
Here  $\gamma_R$ and   $\gamma_D$ are the Rashba and Dresselhaus spin-orbit coefficients. In Fig.~\ref{fig1}, we have plotted the contribution of the Rashba-Dresselhaus spin-orbit coupling ($\alpha_R/\alpha_D$) with the variation of applied electric fields ($E$) along the z-direction. It can be seen that the Rashba spin-orbit coupling dominates in InAs and InSb QDs whereas the Dresselhaus spin-orbit coupling dominates in GaAs and GaSb QDs. In section~\ref{results-and-discussions} we will  focus our investigation on the phonon mediated spin-relaxation  in both symmetric and asymmetric QDs.

The  Hamiltonian~(\ref{rashba-dresselhaus})  can be written in terms of raising and lowering operators as
\begin{widetext}
\begin{eqnarray}
H_{so}&=&\alpha_R \left(1+i\right)\left[
b^{1/4}\kappa_+\left(s_+-i\right)a_++b^{1/4}\kappa_+\left(s_-+i\right)a_-+a^{1/4}\eta_-\left(i-s_-\right)a_++a^{1/4}\eta_-\left(i+s_+\right)a_-
\right]\nonumber\\
&&+\alpha_D
\left(1+i\right)\left[a^{1/4}\kappa_-\left(i-s_-\right)a_++a^{1/4}\kappa_-\left(i+s_+\right)a_-+b^{1/4}\eta_+\left(-i+s_+\right)a_++b^{1/4}\eta_+\left(i+s_-\right)a_-\right]+H.c.,~~~~~~
\label{H-R}
\end{eqnarray}
\end{widetext}
where
\begin{eqnarray}
s_{\pm}=\frac{\omega_+}{\omega_c \left(\frac{b}{a}\right)^{\frac{1}{4}}}\left[\sqrt \frac{b}{a}-1 \pm \left[\frac{\omega_c^2\sqrt \frac{b}{a}}{\omega_+^2}+\left(1-\sqrt \frac{b}{a}\right)^2\right]^{\frac{1}{2}}\right],~~~\\
\kappa_{\pm}=\frac{1}{2\left(s_+-s_-\right)}\left\{\frac{1}{\ell}\sigma_x\pm i\frac{eB\ell}{\hbar}\left(\frac{1}{\sqrt{a}+\sqrt{b}}\right)\sigma_y\right\},~~~~\\
\eta_{\pm}=\frac{1}{2\left(s_+-s_-\right)}\left\{\frac{1}{\ell}\sigma_y\pm i\frac{eB\ell}{\hbar}\left(\frac{1}{\sqrt{a}+\sqrt{b}}\right)\sigma_x\right\},~~~~
\end{eqnarray}
H.c. represents the Hermitian conjugate, $\ell=\sqrt{\hbar/m\Omega}$ is the hybrid orbital length and $\Omega=\sqrt{\omega_0^2+\omega_c^2/\left(\sqrt a + \sqrt b\right)^2}$.

At low electric fields and small QDs radii, we treat
the Hamiltonian associated with the Rashba and linear
Dresselhaus spin-orbit couplings as a perturbation. Using
second order non degenerate perturbation theory, the energy spectrum
of the two lowest electron spin states in QDs (for details, see Ref.~\onlinecite{prabhakar11}) is given by
\begin{eqnarray}
\varepsilon_{0,0,+}=\hbar \varpi_+ - \frac{\alpha_R^2\xi_++\alpha_D^2\varsigma_+}{\hbar \omega_x-\Delta}-\frac{\alpha_R^2\varsigma_-+\alpha_D^2\xi_-}{\hbar \omega_y-\Delta},\\
\varepsilon_{0,0,-}=\hbar \varpi_- - \frac{\alpha_R^2\varsigma_++\alpha_D^2\xi_+}{\hbar \omega_x+\Delta}-\frac{\alpha_R^2\xi_-+\alpha_D^2\varsigma_-}{\hbar \omega_y+\Delta},
\end{eqnarray}
where $\varpi_{\pm}=\omega_+\pm\omega_z/2$, $\omega_z=\Delta/\hbar$ is the Zeeman frequency,  $\omega_x=\omega_++\omega_-$, and $\omega_y=\omega_+-\omega_-$. Also,
\begin{eqnarray}
\xi_{\pm}=\frac{1}{2(s_+-s_-)}\left\{\pm\frac{1}{s_{\pm}}\alpha^2_{\pm}+2\alpha_{\pm}\beta_{\pm} \mp\frac{1}{s_{\mp}}\beta^2_{\pm}\right\},\\
\varsigma_{\pm}=\frac{1}{2(s_+-s_-)}\left\{\pm\frac{1}{s_{\pm}}\alpha^2_{\mp}-2\alpha_{\mp}\beta_{\mp} \mp\frac{1}{s_{\mp}}\beta^2_{\mp}\right\},\\
\alpha_{\pm}=a^{1/4}\left\{\frac{1}{\ell}\pm \frac{eB\ell}{\hbar}\frac{1}{\left(\sqrt{a}+\sqrt{b}\right)}\right\},\\
\beta_{\pm}=b^{1/4}\left\{\frac{1}{\ell}\pm \frac{eB\ell}{\hbar}\frac{1}{\left(\sqrt{a}+\sqrt{b}\right)}\right\}.
\end{eqnarray}

We now turn to the calculation of the phonon induced
spin relaxation rate at absolute zero temperature between two lowest energy states in
QDs. Following Ref.~\onlinecite{prabhakar12,sousa03,khaetskii00,khaetskii01}, the interaction between electron
and piezo-phonon can be written~\cite{gantmakher-book}
\begin{equation}
u^{\mathbf{q}\alpha}_{ph}\left(\mathbf{r},t\right)=\sqrt{\frac{\hbar}{2\rho V \omega_{\mathbf{q}\alpha}}} e^{i\left(\mathbf{q\cdot r} -\omega_{q\alpha} t\right)}e A_{\mathbf{q}\alpha}b^{\dag}_{\mathbf{q}\alpha} + H.c.,
\label{u}
\end{equation}
where  $\rho$ is the crystal mass density and $V$ is the volume of the QD.  $b^{\dag}_{\mathbf{q}\alpha}$ creates an acoustic phonon with wave vector $\mathbf{q}$ and polarization $\hat{e}_\alpha$, where $\alpha=l,t_1,t_2$ are chosen as one longitudinal  and two transverse modes of the induced phonon  in the dots.   $A_{\mathbf{q}\alpha}=\hat{q}_i\hat{q}_k e\beta_{ijk} e^j_{\mathbf{q}\alpha}$ is the amplitude of the electric field created by phonon strain, where $\hat{\mathbf{q}}=\mathbf{q}/q$ and $e\beta_{ijk}=eh_{14}$ for $i\neq k, i\neq j, j\neq k$. The polarization directions of the induced phonon are $\hat{e}_l=\left(\sin\theta \cos\phi, \sin\theta \sin\phi, \cos\theta \right)$, $\hat{e}_{t_1}=\left(\cos\theta \cos\phi, \cos\theta \sin\phi, -\sin\theta \right)$ and $\hat{e}_{t_2}=\left(-\sin\phi, \cos\phi, 0 \right)$. Based on the Fermi Golden Rule, the phonon induced spin transition rate in the QDs is given by~\cite{sousa03,khaetskii01}
\begin{equation}
\frac{1}{T_1}=\frac{2\pi}{\hbar}\int \frac{d^3\mathbf{q}}{\left(2\pi\right)^3}\sum_{\alpha=l,t}\arrowvert M\left(\mathbf{q}\alpha\right)\arrowvert^2\delta\left(\hbar s_\alpha \mathbf{q}-\varepsilon_{f}+\varepsilon_{i}\right),
\label{1-T1}
\end{equation}
where  $s_l$,$s_t$ are the longitudinal and transverse acoustic phonon velocities in QDs.  The matrix element $M\left(\mathbf{q}\alpha\right)=\langle \psi_i|u^{\mathbf{q}\alpha}_{ph}\left(\mathbf{r},t\right)|\psi_f\rangle$   with the emission of one phonon $\mathbf{q}\alpha$ has been calculated perturbatively and numerically.~\cite{khaetskii01,stano06,comsol} Here $|\psi_i\rangle$ and $|\psi_f\rangle$ correspond to the initial and finial states of the Hamiltonian $H$. Based on second order non degenerate perturbation theory, after long algebraic tranformations, we have:
\begin{equation}
\frac{1}{T_1}=c\left(|M_x|^2+|M_y|^2\right),
\label{1-T1-a}
\end{equation}
where
\begin{eqnarray}
c=\frac{2\left(eh_{14}\right)^2\left(g\mu_BB\right)^3}{35\pi \hbar^4\rho}\left(\frac{1}{s^5_l}+\frac{4}{3}\frac{1}{s^5_t}\right),~~~~~~~\label{c}\\
M_x=\frac{\left(is_-+1\right)\Xi_1\left(\hbar\omega_x+\Delta\right)+\left(-is_-+1\right)\Xi_3\left(\hbar\omega_x-\Delta\right)}
{a^{1/4}\left[\left(\hbar\omega_x\right)^2-\Delta^2\right]}\nonumber\\
+\frac{\left(-is_++1\right)\Xi_2\left(\hbar\omega_y+\Delta\right)+\left(is_++1\right)\Xi_4\left(\hbar\omega_y-\Delta\right)}
{a^{1/4}\left[\left(\hbar\omega_y\right)^2-\Delta^2\right]},~~~~~~~\label{M-x}\\
M_y=\frac{\left(is_++1\right)\Xi_1\left(\hbar\omega_x+\Delta\right)+\left(-is_++1\right)\Xi_3\left(\hbar\omega_x-\Delta\right)}
{b^{1/4}\left[\left(\hbar\omega_x\right)^2-\Delta^2\right]}\nonumber\\
+\frac{\left(is_--1\right)\Xi_2\left(\hbar\omega_y+\Delta\right)+\left(-is_--1\right)\Xi_4\left(\hbar\omega_y-\Delta\right)}
{b^{1/4}\left[\left(\hbar\omega_y\right)^2-\Delta^2\right]},~~~~~~\label{M-y}
\end{eqnarray}
\begin{widetext}
\begin{eqnarray}
\Xi_1=\frac{\ell}{2\left(s_+-s_-\right)^2}\left[\alpha_R\left\{\left(s_++i\right)\beta_++\left(1-is_-\right)\alpha_+\right\}
+\alpha_D\left\{\left(-s_--i\right)\alpha_-+\left(-1+is_+\right)\beta_-\right\}\right],~~~~\label{Xi-1}\\
\Xi_2=\frac{\ell}{2\left(s_+-s_-\right)^2}\left[\alpha_R\left\{\left(s_--i\right)\beta_++\left(1+is_+\right)\alpha_+\right\}
+\alpha_D\left\{\left(s_+-i\right)\alpha_-+\left(1+is_-\right)\beta_-\right\}\right],~~~~\label{Xi-2}\\
\Xi_3=\frac{\ell}{2\left(s_+-s_-\right)^2}\left[\alpha_R\left\{\left(s_+-i\right)\beta_-+\left(-1-is_-\right)\alpha_-\right\}
+\alpha_D\left\{\left(-s_-+i\right)\alpha_++\left(1+is_+\right)\beta_+\right\}\right],~~~~\label{Xi-3}\\
\Xi_4=\frac{\ell}{2\left(s_+-s_-\right)^2}\left[\alpha_R\left\{\left(s_-+i\right)\beta_-+\left(-1+is_+\right)\alpha_-\right\}
+\alpha_D\left\{\left(s_++i\right)\alpha_++\left(-1+is_-\right)\beta_+\right\}\right].~~~~\label{Xi-4}
\end{eqnarray}
\end{widetext}
In the above expression, we use $c=c_l I_{xl}+2c_tI_{xt}$, where $c_\alpha=\frac{q^2e^2}{\left(2\pi\right)^2\hbar^2s_\alpha}|\varepsilon_{q\alpha}|^2$, $|\varepsilon_{q\alpha}|^2=\frac{q^2\hbar}{2\rho\omega_{q\alpha}}$ and $q=\frac{g\mu_BB}{\hbar s_\alpha}$. Also, $g=\frac{\varepsilon_{0,0,-}-\varepsilon_{0,0,+}}{\mu_BB}$ is the Land$\acute{e}$ $g$-factor. For longitudinal phonon modes,~\cite{khaetskii01,golovach04} we have $|A_{q,l}|^2=36h_{14}^2 \cos^2\theta\sin^4\theta\sin^2\phi\cos^2\phi$ and thus we find $I_{xl}=16\pi h^2_{14}/35$.
For transverse phonon modes, we have
$|A_{q,t}|^2=2h_{14}^2\left[\cos^2\theta\sin^2\theta + \sin^4\theta\left(1-9\cos^2\theta\right)\sin^2\phi\cos^2\phi\right]$ and thus we find $I_{xt}=32\pi h^2_{14}/105$.

For isotropic QDs ($a=b=1$, $s_+=1$ and  $s_-=-1$), the spin relaxation rate is given by
\begin{equation}
\frac{1}{T_1}=\frac{2\left(eh_{14}\right)^2\left(g\mu_BB\right)^3}{35\pi \hbar^4\rho}\left(\frac{1}{s^5_l}+\frac{4}{3}\frac{1}{s^5_t}\right)\left(|M_R|^2+|M_D|^2\right),
\label{1-T1-1}
\end{equation}
where $M_R$ and $M_D$ are the coefficients of matrix elements
associated with the Rashba and Dresselhaus spin-orbit couplings
in QDs and are given by
\begin{eqnarray}
M_R=\frac{\alpha_R}{\sqrt 2 \hbar \Omega}\left[\frac{1}{1-\frac{\Delta}{\hbar\left(\Omega+\frac{\omega_c}{2}\right)}}-
\frac{1}{1+\frac{\Delta}{\hbar\left(\Omega-\frac{\omega_c}{2}\right)}}\right],\label{M-R}\\
M_D=\frac{\alpha_D}{\sqrt 2 \hbar \Omega}\left[\frac{1}{1+\frac{\Delta}{\hbar\left(\Omega+\frac{\omega_c}{2}\right)}}-
\frac{1}{1-\frac{\Delta}{\hbar\left(\Omega-\frac{\omega_c}{2}\right)}}\right].\label{M-D}
\end{eqnarray}
Since $\Delta = g_0\mu_BB$ is negative for GaAs and InAs QDs, we see the degeneracy only appears in the Rashba case
(see the 2nd term of Eq.~\ref{M-R}) and the degeneracy is absent in the Dresselhaus case.  The degeneracy in the Rashba case induces the
level crossing point and cusp-like structure in the spin-flip rate in QDs. The
spin relaxation rate for isotropic QDs can be written in a more convenient form as
\begin{eqnarray}
\frac{1}{T_1}&=&\frac{2\left(eh_{14}\right)^2\left(g\mu_BB\right)^3}{35\pi \hbar^4\rho}\left(\frac{1}{s^5_l}+\frac{4}{3}\frac{1}{s^5_t}\right)\frac{2\Delta^2 m^4}{\hbar^8} \nonumber\\ && \left(\alpha_R^2+\alpha_D^2\right) \ell_0^8
\left[1+O \left(\omega_c/\omega_o\right)^2\right].
\label{1-T1-2}
\end{eqnarray}
From  Eq.~\ref{1-T1-2}, it is clear that the spin-flip rate vanishes like $B^5$ and $\ell_0^8$ (see Ref.~\onlinecite{sousa03}).

\begin{table}[b]
\caption{\label{table1}%
The material constants used in our calculations are taken from Refs.~\onlinecite{cardona88,sousa03}
}
\begin{ruledtabular}
\begin{tabular}{llcdr}
Parameters & GaAs & InAs & GaSb & InSb \\
\colrule
$g_0$ &-0.44  & -15 & -7.8 & -50.6 \\
m & 0.067 & 0.0239 & 0.0412& 0.0136\\
$\gamma_R~[{\AA}^2]$ & 4.4 & 110 & 33& 500\\
$\gamma_D~[eV{\AA}^3]$ & 26 &130 & 187&228\\
$eh_{14}~[10^{-5}erg/cm]$ & 2.34 &0.54 & 1.5 & 0.75\\
$s_l~[10^{5}cm/s]$ & 5.14 &4.2 & 4.3 & 3.69\\
$s_t~[10^{5}cm/s]$ & 3.03 &2.35 & 2.49 & 2.29\\
$\rho~[g/cm^3]$ & 5.3176 &5.667 & 5.6137 & 5.7747\\
\end{tabular}
\end{ruledtabular}
\end{table}

\section{Computational Method}\label{computational-method}

We suppose that a  QD is formed at the center of a $400\times 400~\mathrm{nm^2}$ geometry.  Then we diagonalize the total Hamiltonian  $H$ numerically using the Finite Element Method.~\cite{comsol} The geometry contains  $24910$ elements. Since the geometry is much larger compared to the actual lateral size of the QD,  we impose Dirichlet boundary conditions, find the  eigenvalues, eigenfunctions and the  matrix elements $M\left(\mathbf{q}\alpha\right)$ of the total Hamiltonian $H$. From  Figs.~\ref{fig2} to ~\ref{fig7}, the analytically obtained spin-flip rates from Eq.~\ref{1-T1-a} (solid and dashed-dotted lines) are seen to be in excellent agreement with the numerical values (open circles and squares). The material constants are taken from table~\ref{table1}.

\section{Results and Discussions}\label{results-and-discussions}

In Fig.~\ref{fig2}, we compare theoretically obtained spin-flip rates from Eq.~\ref{1-T1-a} to the experimentally reported values in Ref.~\onlinecite{amasha08}. Theoretical and experimental data are in excellent agreement. Inset plots (from left to right) show  realistic in-plane wavefunctions of  QDs for the spin states $|0,0,+1/2\rangle$, $|0,0,-1/2\rangle$ and $|0,1,+1/2\rangle$. It can be seen that anisotropy breaks the in-plane rotational symmetry. As a result, we find that the in-plane wavefunction of anisotropic QDs for the states $|0,1,+1/2\rangle$  split into two which has a direct consequence on inducing accidental degeneracy even for the pure  Dresselhaus spin-orbit coupling case in the phonon mediated spin flip rate. This  will be separately discussed from Figs.~\ref{fig3} to~\ref{fig7}.

In Fig.~\ref{fig3} (i) we see that the cusp-like structure is absent (i.e., the spin-flip rate is a monotonous function of the magnetic field) for the pure Dresselhaus case in symmetric QDs. However in Fig.~\ref{fig3}(iii) we see that the cusp-like structure is present for the pure Dresselhaus case in asymmetric QDs.
In Fig.~\ref{fig4}, again we see that the cusp-like structure is absent in isotropic QDs ($a=b$) but is present in anisotropic QDs ($a\neq b$) for the pure Dresselhaus case. The cusp-like structure  in anisotropic QDs is thus due to the fact that the anisotropy induces the accidental degeneracy in the matrix elements  ($M\left(\mathbf{q}\alpha\right)$) near the level crossing or anticrossing point. The accidental degeneracy point where the cusp-like structure appears  is referred to as the spin-hot spot while tuning on  the spin-orbit coupling removes the degeneracy.~\cite{stano06} Thus, we apply degenerate perturbation theory  and the energy spectrum of the  unperturbed spin states $|0,0,-\rangle$ and $|0,1,+\rangle$ for anisotropic QDs are given by
\begin{eqnarray}
\varepsilon^0_{0,0,-}=\frac{3}{2}\hbar \omega_+ -\frac{1}{2}\hbar \omega_-+\left[\alpha_R^2\xi_-+\alpha_D^2\zeta_-\right]^{1/2},\label{correction_term-1}\\
\varepsilon^0_{0,1,+}=\frac{3}{2}\hbar \omega_+ -\frac{1}{2}\hbar \omega_--\left[\alpha_R^2\xi_-+\alpha_D^2\zeta_-\right]^{1/2}.\label{correction_term-2}
\end{eqnarray}
We have substituted Eqs.~\ref{correction_term-1} and~\ref{correction_term-2} into \ref{1-T1-a} and  found  the spin-flip rate at  the level crossing point from Figs.~\ref{fig2} to~\ref{fig7}.
Lifting the degeneracy with the application of spin-orbit couplings
mixes spin up and spin down states where the phonon mediated spin transition rate
between states of opposite magnetic moment will involve
spin flips with a much more enhanced probability compared
to the normal states. For example, the spin-hot spot for the pure Dresselhaus case in symmetric GaAs QDs (Figs.~\ref{fig3} (i) and ~\ref{fig4} (i)) can not be observed while tuning the anisotropy ($a\neq b$), however can be observed  at $B=5.1~T$ and $\ell_0=69~nm$ as shown in Figs.~\ref{fig3} (iii) and ~\ref{fig4} (iii), respectively. Notice that the spin-flip rates of the pure Dresselhaus case  found near the spin-hot spot in Figs.~\ref{fig3} (iii) and ~\ref{fig4} (iii) are $6$ orders of magnitude larger than  those  values found in Figs.~\ref{fig3} (i) and~\ref{fig4}(i). This result (i.e., the spin-hot spot in asymmetric QDs for the pure Dresselhaus case yet to be experimentally verified)  provides small relaxation  and  decoherence time which should be avoided during the design of  spin based transistors for the possible implementation in quantum logic gates, quantum computing and quantum information processing.
From Figs.~\ref{fig4} to~\ref{fig7}, we investigated the spin relaxation rate in InAs, GaSb and InSb QDs. Analyzing  all plots,  the spin-hot spot and associated cusp-like structure can  be seen in the pure  Dresselhaus spin-orbit coupling case in anisotropic QDs ($a\neq b$).

\section{Conclusions}\label{conclusion}

We have shown that the anisotropy breaks the in-plane rotational symmetry. As a result, we found that the cusp-like structure (i.e., where the spin-hot spot) is present  in the phonon mediated spin transition rate in anisotropic  QDs for the pure Dresselhaus spin-orbit coupling case. In contrast, for isotropic QDs,  the spin transition rate is a monotonous function of magnetic fields and QDs radii (i.e., where the spin-hot spot is absent) for the pure Dresselhaus spin-orbit coupling case. These  results (yet to be experimentally verified) provide new information for finding the spin hot-spot in anisotropic spin relaxation   for the pure Dresselhaus case during the  design of  QD spin transistors. At or nearby the spin-hot spot, the relaxation and  decoherence time are smaller by several orders of magnitude. One should  avoid such  locations during the design of QD spin based transistors for the possible implementation in quantum logic gates, quantum computing and quantum information processing.

\section{Acknowledgements}\label{acknowlwdgement}
This work has been supported by Natural Science and Engineering Research Council (Canada) and Canada Research Chair programs. The authors acknowledge the Shared Hierarchical Academic Research Computing Network (SHARCNET) community  and Dr. Philip James Douglas Roberts for the helpful and technical support.


%

\end{document}